\documentstyle[epsfig,times]{l-aa}

\def\ltsima{$\; \buildrel < \over \sim \;$}
\def\simlt{\lower.5ex\hbox{\ltsima}}
\def\gtsima{$\; \buildrel > \over \sim \;$}
\def\simgt{\lower.5ex\hbox{\gtsima}}

\begin{document}
%\large
   \thesaurus{13.25.2 -- 11.19.1 -- 10.09.1}
   \title{Hard X--ray detection of NGC 1068 with BeppoSAX}

   \author{G. Matt 
          \inst{1}
   \and M. Guainazzi
          \inst{2}
   \and F. Frontera 
          \inst{3} 
   \and L. Bassani
          \inst{3}
   \and W.N. Brandt
  	  \inst{4}
   \and A.C. Fabian
  	  \inst{5}
   \and F. Fiore
          \inst{2,6} 
   \and F. Haardt	
          \inst{7}
   \and K. Iwasawa	
          \inst{5}
   \and R. Maiolino
          \inst{8}
   \and G. Malaguti	
          \inst{3}
   \and A. Marconi	
	  \inst{9,10}
   \and A. Matteuzzi	
          \inst{2}
   \and S. Molendi
	  \inst{11}	
   \and G.C. Perola	
          \inst{1}
   \and S. Piraino
          \inst{12}
   \and L. Piro	
	  \inst{13}
}

   \offprints{G. Matt (matt@amaldi.fis.uniroma3.it)}

%1
   \institute{Dipartimento di Fisica ``E. Amaldi", 
              Universit\`a degli Studi ``Roma Tre", 
              Via della Vasca Navale 84, I--00146 Roma, Italy
%2
   \and SAX/SDC Nuova Telespazio, Via Corcolle 19,  I--00131 Roma, Italy
%3
   \and Istituto Tecnologie e Studio Radiazioni Extraterrestri, CNR, Via
              Gobetti 101, I--40129 Bologna, Italy
%4 
    \and Harvard--Smithsonian Center for Astrophysics, 60 Garden St.,
	Cambridge, MA 02138, USA
%5
     \and Institute of Astronomy,
	University of Cambridge, Madingley Road, Cambridge CB3 0HA, U.K.
%6
      \and Osservatorio Astronomico di Roma, Via dell'Osservatorio,
	I--00044 Monteporzio-Catone, Italy
%7
     \and Dipartimento di Fisica, Universit\`a di Milano, Via Celoria 16,
       I--20133 Milano, Italy
%8
   \and MPI f\"ur Extraterrestrische Physik, Giessenbachstrasse 1,
             D--85748 Garching bei M\"unchen, Germany
%9
   \and Space Telescope Science Institute, 3700 San Martin Drive, Baltimore, 
	MD 21218, U.S.A.
%10
    \and Dipartimento di Astronomia e Scienza dello Spazio,
	Largo E. Fermi 5, I--50125 Firenze, Italy
%11
   \and Istituto di Fisica Cosmica e Tecnologie Relative, Via Bassini 15, 
	I--20133 Milano, Italy
%12
   \and Istituto di Fisica Cosmica ed Applicazioni dell'Informatica, Via
             Ugo La Malfa 153, I--90146 Palermo, Italy
%13
   \and Istituto di Astrofisica Spaziale -- C.N.R., Via E. Fermi 21, 
		I--00044 Frascati, Italy	
%    \and Osservatorio Astrofisico
%             di Arcetri, L.~E.Fermi 5, I--50125 Firenze, Italy
%   \and Osservatorio Astronomico di Brera, Via Brera 28, I--20121 
%             Milano, Italy
   }

   \date{Received / Accepted }

   \maketitle

   \begin{abstract}
We report on the first detection above $\sim$10 keV of the archetypal 
Compton--thick Seyfert 2 galaxy NGC 1068. 
This detection, obtained with the PDS instrument onboard BeppoSAX
in the 20--100 keV range, confirms the hardness of the X--ray spectrum
above a few keV (as indicated by ASCA observations) and 
supports models envisaging a mixture 
of both neutral and ionized reflections of an otherwise invisible nuclear
continuum, the neutral reflection component being the dominant one in hard
X--rays.

      \keywords{X-rays: galaxies -- Galaxies: Seyfert -- Galaxies: individual: 
       NGC1068}
   \end{abstract}

%
%  14.Sep.'90: Demo-Vs.
%________________________________________________________________

\section{Introduction}

NGC~1068 is the archetypal object for both the class of Seyfert 2 galaxies 
as a whole and the Compton--thick subclass (see Matt 1997 for a 
brief review), i.e. those
objects for which the column density of the 
line--of--sight absorbing matter (hereinafter
identified with the torus, see e.g. Ward 1996) exceeds 
$\sigma_{\rm T}^{-1}=1.5\times10^{24}$ cm$^{-2}$
and is therefore optically thick to Compton scattering. A column of a few times
this critical density is sufficient to strongly depress the transmitted 
intensity also in hard X--rays, since after a few scatterings photons are 
redshifted into the photoelectric dominated regime. However, even for
very high column densities, nuclear 
radiation can still be observed in scattered light, as both the visible
part of the inner surface of the torus and the warm medium responsible
for scattering and polarizing the optical broad lines (Antonucci \&
Miller 1985) can act as mirrors (e.g. Ghisellini et al. 1994;
Matt et al. 1996). The two reflectors produce rather
different spectra: the torus is supposed to be essentially neutral,
then giving rise to the so--called Compton reflection continuum
(Lightman \& White 1988); the warm mirror should be highly ionized,
and its spectral shape basically the same as the nuclear one,
apart from a high energy cutoff owing to Compton downscattering.
Superimposed on both continua, emission lines are also expected, iron lines
being the most prominent: a 6.4 keV fluorescent line from the torus, highly
ionized resonant scattering and fluorescence/recombination lines from the 
warm mirror (such as reported by Matt et al. 1996). 

This two--reflectors picture is confirmed (and it has been partly 
motivated) by ASCA observations of NGC~1068 (Ueno et al. 1994; Iwasawa
et al. 1997), which clearly distinguished three different iron
lines: one at 6.4 keV, the other two consistent with He-- 
and H--like iron,
respectively (the lines being possibly redshifted by a few thousands 
km/s, Iwasawa et al. 1997) with
equivalent widths of the order of 1 keV each, as expected in the 
reflection model (Matt et al. 1996). The continuum is not well constrained
by ASCA due to the limited energy range available for its evaluation
(below 4 keV a thermal--like component dominates, Ueno et al. 1994),
but it looks very flat, as expected if the cold reflector were providing a
significant contribution.

To confirm this scenario,  
 hard X--rays observations are necessary. We then
proposed to observe NGC~1068 with BeppoSAX, in
order to take full advantage of the good sensitivity of the PDS, the Phoswich
Detector System working between 15 and 300 keV. In fact, a cold 
reflection--dominated spectrum would be hard enough to be observable 
with the PDS with a $\sim$10$^5$ s pointing 
even at the relatively low flux level of NGC~1068 (i.e. about a quarter of a
mCrab\footnote{1 mCrab=2.4$\times 10^{-11}$ and 3$\times 10^{-11}$ erg cm$^{-2}$
s$^{-1}$ in the 2--10 and 20--200 keV ranges, respectively.}
  between 2 and 10 keV). 
Here we report what is to our knowledge
the first  detection of NGC 1068 up to 100 keV.

\section{Data reduction}

The X-ray satellite BeppoSAX (Boella et al. 1997a), a program of the 
Italian space agency (ASI)
with participation of the Netherlands agency for Aereospace Program (NIVR),
includes 
four co--aligned Narrow Field Instruments: a Low Energy
Concentrator Spectrometer (LECS), three 
Medium Energy Concentrator Spectrometers
(MECS), a High Pressure Gas Scintillation Proportional 
Counter (HPGSPC), and a Phoswich Detector System (PDS).
The LECS and MECS have 
imaging capabilities and cover the 0.1--10 keV and 1.3--10 keV energy ranges
respectively; in the overlapping band the total effective area of the
MECS (which is $\sim$150 cm$^2$ at 6 keV)
is about three times that of the LECS.
The energy resolution is  $\sim$8\% and the angular resolution is 
$\sim$1.2 arcmin (Half Power radius) at 6~keV for both instruments. 
The HPGSPC and the PDS are collimated instruments covering the 
4--120 keV and the 15--300 keV energy ranges respectively. In the overlapping
band the PDS is more sensitive, the best characteristic of the HPGSPC
being instead the good energy resolution.  

BeppoSAX observed the source from December 30 1996 to January 3 1997
for about 110 ks effective time. In this paper we present 
only MECS (Boella et al. 1997b) and PDS (Frontera et al. 1997) data,
because of our choice to restrict the analysis to energies greater than
4 keV (see next section).

MECS spectra have been extracted from a 4 arcmin radius region around the
centroid of the source image; the 
spectra from the three units have been equalized to the
MECS1 energy--PI relationship and added together. Since the MECS 
background is very low and stable\footnote{For information on the 
background and on data analysis in general see
{\sl http://www.sdc.asi.it/software/cookbook} }, 
data selection is straightforward; we used
all data acquired with an angle, with respect to the Earth limb,
higher than 5$^{\circ}$. The background subtraction has been performed using
blank sky spectra extracted from the same region of the detector
field of view. 

The PDS consists of four units, and was operated 
in collimator rocking mode, with a pair of 
units pointings to the source and the other pair pointings $\pm 210$ arcmin 
away, the two pairs switching on and off source every 96 seconds. 
The net source spectra have
been obtained by subtracting the `off' to the `on' counts.
Then the spectra of the four crystals have been summed together after
performing gain equalization. A slight improvement in the 
S/N has been obtained by optimizing the Rise Time selection:
only events in the range 5--133 (digital units; 
see Fig. 2 of Frontera et al. 1997) have been used. This choice guarantees no
reduction in the detection efficiency due to pulse shape analysis. 
Isolated spikes, likely due to short particle ``bursts'',
 have been removed. Data within 5 minutes after any 
South Atlantic Anomaly passage have
 been discarded to avoid gain problems due to high voltage recovery
 after instrument switch--off. The total effective on--source
time is T$_{\rm exp}$=65000 s.
The net count rate (CR) is $0.15\pm0.03$ cts/s,
corresponding to a 20--200 keV flux $\sim$1.3 mCrab if a simple power--law
with $\Gamma$=2 is assumed (note that this flux is
well below the OSSE and SIGMA upper limits, McNaron--Brown et al., in 
preparation,  Jourdain et al. 1994). 

As a further check, we have divided the whole observation in three
temporal segments; the net count rates are all consistent one another: 
CR$_{first} = 0.15 \pm 0.05$ cts/s (T$_{\rm exp} \simeq$  22400 s), CR$_{second}
= 0.13 \pm 0.05$ cts/s (T$_{\rm exp} \simeq$ 20300 s), CR$_{third} = 0.15 \pm 0.05$
cts/s (T$_{\rm exp} \simeq$ 21800 s).

Deep observations of blank fields show that the systematic residuals 
in the background subtraction procedure are at most 0.25 mCrab in the
20--200 keV energy range where the PDS response matrix is
currently well calibrated.
%(Guainazzi \& Matteuzzi, 1997). 
Even if this maximum 
value were subtracted from the observed count rate,
the S/N would remain $> 4 \sigma$.

We have also checked for the presence of contaminating 
sources in the $\simeq 1.3^{\circ}$ PDS field of view. No source
at a flux level $>$0.03 mCrab in the 2--10 keV band is reported
in X--ray catalogs. In particular, no source is present neither in
the PDS field of view nor in an annulus at angle $\theta \simeq 210$ arcmin
from the pointing position in the hard X--ray extragalactic sources catalog
of Malizia \& Bassani (1996).

\section{Spectral analysis and results}

Spectral fits have been performed with the {\sc xspec 9.0} package,
using the response matrices released on Jan 1997. 
To cure the mismatch between MECS and PDS
absolute normalizations in the current matrices,
PDS data have been divided by a factor 0.7, constant over
energy (Cusumano et al., in preparation). 
%(In practice,
%any fitting model has been multiplied by a constant factor of 0.7 when applied
%to the PDS data).  
Note that even allowing 
for a 10\% remaining uncertainty in the cross--calibration, the continuum
best fit parameters 
would change only slightly, and the basic picture would by no means be altered. 

In the following, all quoted errors correspond 
to 90\% confidence level for one interesting parameter ($\Delta\chi^2$=2.7).

As we are interested here only in the high energy part of the spectrum (i.e.
that believed to be due to the reflection of the obscured nucleus), we
restricted our analysis to energies greater than 4 keV to avoid contamination
from other spectral components. 
The overall spectrum between 4 and 100 keV
is shown in Fig.1. The spectrum confirms that
NGC~1068 is substantially Compton--thick, otherwise it would 
have been detected at a much higher
level in hard X--rays, similarly to NGC~4945 (Iwasawa et al. 1993; 
Done et al. 1996). A prominent, broad iron line is clearly seen.
According to ASCA--SIS results (Ueno et al. 1994; Iwasawa et al. 1997),
we fitted this broad feature with a blend of three narrow lines, 
with energies fixed at 6.4 keV (corresponding to neutral iron), 
6.7 keV (He--like
iron) and 6.97 keV (H--like iron), respectively. The results are
presented in Table 1, and have been obtained adopting a power law
for the continuum (fitted to MECS data only).
No significant differences in the parameters of the lines
have been found with a more complex
description of the continuum (see below). If the ionized lines energies are 
instead fixed at 6.61 and 6.86 keV, as suggested by Iwasawa et al. (1997),
the results are somewhat different, the H--like line being now more intense
than the He--like one. Leaving these energies free, the
best fit values are 6.64 and 7.01 respectively (but consistent within the 
errors with the nominal atomic values). From a statistical point of view, the
first and third fits seem to be preferred, but all three results are 
acceptable at the 90\% confidence level. 
The power law index of the underlying continuum is insensitive
to the details of the line modeling. What is important to remark
is that the lines' equivalent widths are of the expected order if these lines
were produced by reflection of an invisible primary radiation (Matt 
et al. 1996). 

If the MECS+PDS continuum is fitted with a simple power law, a good
fit is obtained (see Table 2). (The iron lines have been modeled for simplicity
with a single, broad gaussian feature).
However, fitting the PDS data alone
gives a photon index of 1.83$^{+0.72}_{-0.52}$, inconsistent at the 
90\% level with the index of the total spectrum.  
This is due to the fact that 
the index is driven basically by the MECS, owing to its better
statistics, while the PDS data, even if lying, on average, on the
extrapolation of the lower energy spectrum, have a different spectral
shape. This high energy steepening, together with 
the flatness of the 4--10 keV continuum
and the presence of a strong 6.4 keV iron line 
suggests that part of the continuum is due to reflection 
of the nuclear radiation by circumnuclear neutral matter, possibly the 
inner surface of the torus.
The presence of intense He-- and H--like iron lines, and the fact that
the 4--10 keV continuum, even if flatter than usual in Seyfert galaxies, 
is not as flat as
a pure cold reflection spectrum would be, indicates contribution from an 
ionized reflector too, to be identified
with the same medium responsible for the scattering and polarization of
the optical broad lines. We then fitted the MECS+PDS continuum with a cold 
reflection component
({\sc xspec}'s model {\sc plrefl}) plus reflection from free electrons,
i.e. a power law with Compton downscattering in the assumption
of a parallel incident beam and temperature of the electrons negligible
with respect to the photon energy; in formulae
(e.g. Matt 1996 and Poutanen et al. 1996):

\begin{equation}
I(E) = {3 \sigma_{\rm T} \over 16\pi}
 \left({d\Omega \over 4\pi}\right) 
\left[ {E \over E_0} + {E_0 \over E} - \sin^2\theta \right] 
I(E_0) ~~,
\end{equation}

\noindent
where $\theta$ is the scattering angle, $d\Omega$ the
solid angle subtended by the illuminated matter, and

\begin{equation}
E_0 = {E \over 1 - \left(E/mc^2\right) (1 - \cos\theta)} ~~.
\end{equation}

\noindent
The power law index of the illuminating radiation has been
assumed to be the same for the two reflection components
(so assuming no angular dependence of the nuclear emission
spectral shape). 
To agree with the current wisdom on Seyfert 1 X--ray spectra,
a reflection component from the accretion
disc should actually have been included in modeling the nuclear radiation;
however, it appears to be an unnecessary sophistication here. 
The fit results are shown in Table 2, and the best fit model in Fig.2. 
The power law index of the primary emission (1.74$^{+0.25}_{-0.56}$)
is consistent with typical Seyfert 1 values (Nandra \& Pounds 1994). The
scattering angle $\theta$ (which, with our assumptions, is the same as
the system inclination angle) is unfortunately not 
constrained, affecting the
spectrum only at the highest energies, where the statistics is
poor. The warm reflection component is the most important
in the 4--10 keV
range, while the cold reflection component dominates above ~10 keV (see Fig.2). 

If the nuclear X--ray luminosity is of the
order of 10$^{44}$ erg s$^{-1}$, as indicated by several and
independent pieces of evidence (Iwasawa et al. 1997 and 
references therein), then the 20--100 keV observed flux is about 2 orders 
of magnitude lower than the nuclear one. Adopting the Ghisellini et al. 
(1994) geometry, {\it both} a very high equatorial 
column density of the torus ($\simgt$10$^{26}$ cm$^{-2}$) (to avoid a
significant transmitted intensity) and a high
inclination angle (to reduce the observable fraction of the inner surface of 
the torus) are implied 
(see Fig.6 of Ghisellini et al. 1994). A high inclination disagrees with
estimates based on both polarization properties (Miller 
et al. 1991), assuming that
the X--ray and optical/near IR scatterers are one and the
same, and infrared mapping of the torus (Young et al. 1996);
such a disagreement would suggest that either the geometry is more 
complex than commonly 
assumed or the luminosity is smaller than estimated. Interestingly, 
however, both
the large column density and the high inclination are consistent with 
recent water maser measurements (Gallimore et al. 1996; Greenhill et al. 1996).

\begin{table}
\caption{Iron lines parameters.
The lines have been fitted with $\delta$--functions, 
the continuum with a simple power law over the 4 to 10.5
keV range (i.e. MECS data only). Line
energies refer to the source rest frame ($z$=0.0038).
Line fluxes are in  10$^{-5}$ ph cm$^{-2}$ s$^{-1}$. }

\vglue0.3truecm
{\hfill\begin{tabular}{l l l l l}
Photon & Line Energy  & Line Flux  & EW  & $\chi^2$/d.o.f. \\
index  & (keV) &   & (keV)  & ($\chi_r^2$) \\
&&\\
\hline  
&&\\
1.17$\pm$0.14 &   6.40~(fixed)  & 3.5$\pm$0.8    & 0.75   & 117/107 \\
              &   6.70~(fixed)  & 6.1$\pm$1.1    & 1.40   & (1.09) \\
              &   6.97~(fixed)  & 4.0$\pm$0.9    & 0.96   & \\
&&\\
\hline
&&\\
1.14$\pm$0.14 &   6.40~(fixed)  & 3.2$\pm$1.0    & 0.69   & 125/107 \\
              &   6.61~(fixed)  & 3.6$\pm$1.5    & 0.79   & (1.17) \\
              &   6.86~(fixed)  & 6.7$\pm$1.0    & 1.55   & \\
&&\\
\hline
&&\\
1.18$\pm$0.15 &   6.40~(fixed)  & 2.4$\pm$0.9   & 0.53   & 111/105 \\
              &   6.64$\pm$0.11 & 7.1$\pm$1.8   & 1.65   & (1.06) \\
              &   7.01$\pm$0.10 & 4.4$\pm$0.9   & 1.07   & \\
&&\\
\hline
\end{tabular}\hfill}
\end{table}

\begin{table}
\caption{MECS+PDS joint fits with either a simple power law or 
the two--reflectors model (see text for details). 
Fluxes are in 10$^{-12}$ erg cm$^{-2}$ s$^{-1}$ and refer
to the continuum only. C and W stand for the
cold and warm reflector respectively.}

\vglue0.3truecm
{\hfill\begin{tabular}{l l l l l }
~ & Photon &  Flux   & Flux  & $\chi^2$/d.o.f. \\
~ &index  & (4-10 keV) & (20--100 keV)  & ($\chi_r^2$) \\
&&\\
\hline
&&\\
Power & 1.14$^{+0.11}_{-0.09}$ & 2.78  & 19.2  & 121/113 \\
law &       &      &       &    (1.07) \\   
&&\\
\hline  
&&\\
Two & 1.74$^{+0.25}_{-0.56}$ & 0.9 (C) & 17.7 (C)  & 118/112 \\
refl. &       &  1.8 (W)    &  3.5 (W)     &   (1.05) \\
&&\\
\hline

\end{tabular}\hfill}
\end{table}

\begin{figure}
\epsfig{file=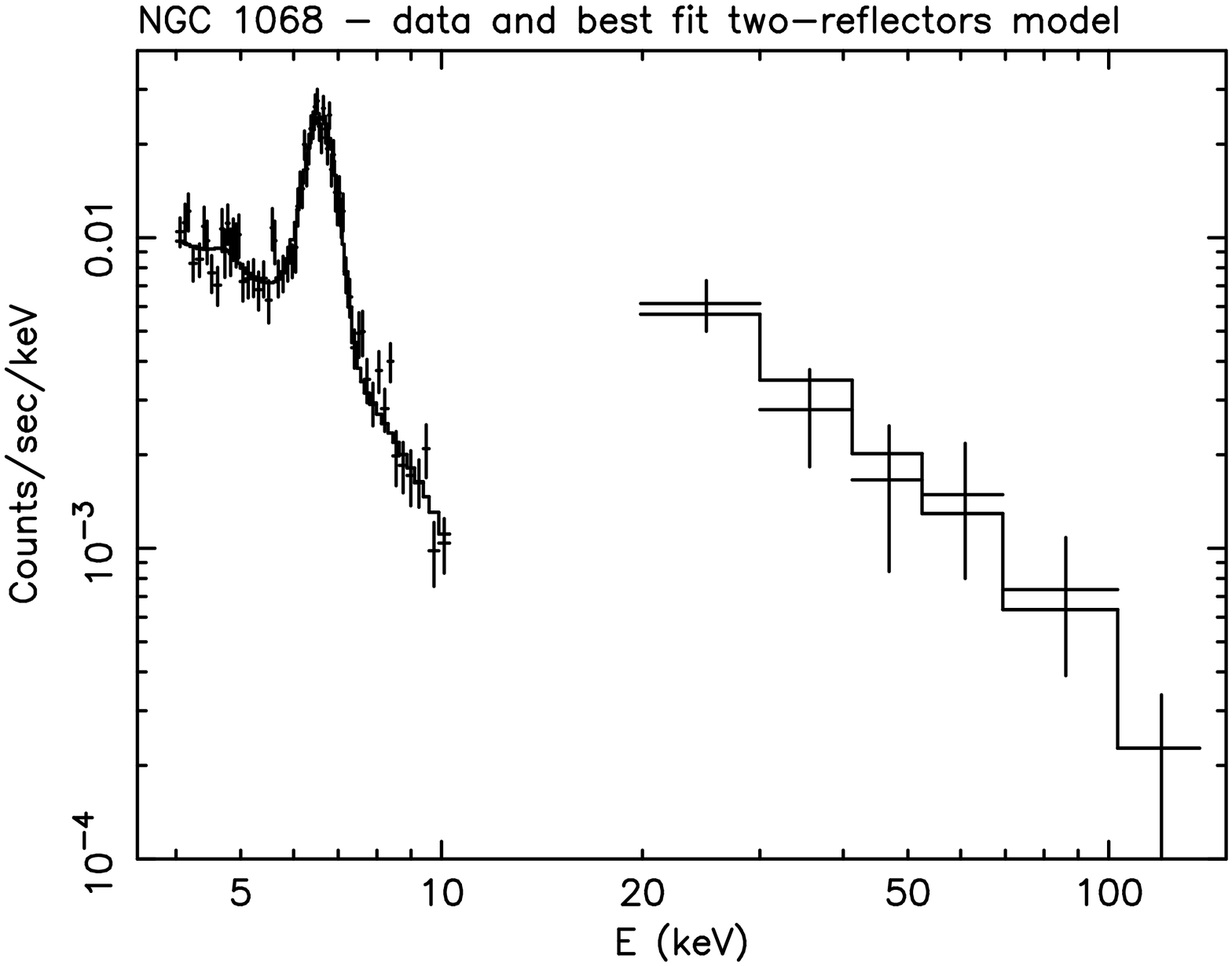, height=9.5cm, angle=-90, silent=}
\caption{The observed BeppoSAX spectrum of NGC~1068 (folded with the
instruments response) above 4 keV. Best fit
parameters (two--reflectors model) are given in Table 2.}
\end{figure}

\begin{figure}
\epsfig{file=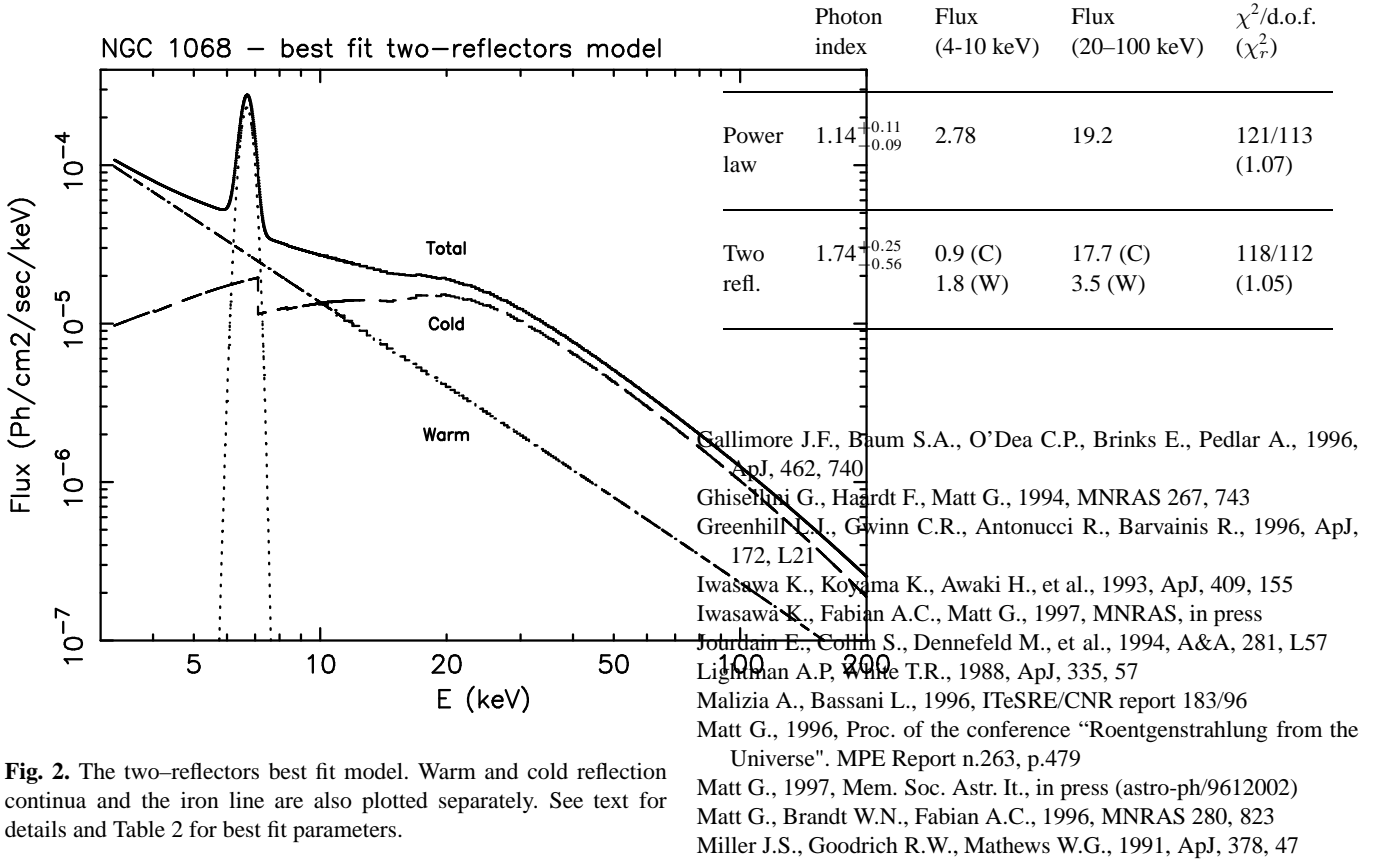, height=9.5cm, angle=-90, silent=}
\caption{The two--reflectors best fit model. Warm and cold reflection
continua and the iron line are also plotted separately.
 See text for details and Table 2 for
best fit parameters.}
\end{figure}

\section{Summary}
We have reported on the detection of the Seyfert 2
galaxy NGC~1068 up to 100 keV, obtained with the PDS instrument onboard 
BeppoSAX. This is the first time that a Compton--thick Seyfert 2 is 
detected at so high energies, supporting the scenario in which 
medium to hard X--rays are due to reflection, from both ionized and
neutral circumnuclear media,
of an otherwise invisible nuclear radiation.

\begin{acknowledgements}
We thank all the people who, at all levels, have made possible the SAX mission.
This research has made use of SAXDAS linearized and cleaned event
files (rev0.1) produced at the BeppoSAX Science Data Center. GM and GCP 
acknowledge financial support from an ASI grant, AMar from
STScI (GO grant G005.44800).
\end{acknowledgements}

\end{document}